\documentclass[
    aps,
    prb,
    showpacs,
    reprint,
    longbibliography,
    superscriptaddress,
    floatfix,
    amsmath
]{revtex4-1}

\pdfoutput=1
\usepackage{braket}
\usepackage{graphics}
\usepackage{graphicx}
\usepackage{multirow}
\usepackage{tasks}
\usepackage{dcolumn}
\usepackage{scrextend}
\usepackage{longtable}
\usepackage{dblfnote}
\usepackage[export]{adjustbox}
\usepackage[multiple]{footmisc}
\newcolumntype{.}{D{.}{.}{-1}}

\usepackage{epsfig}
\usepackage{epstopdf}
\usepackage{hyperref}
\usepackage{verbatim}
\usepackage[english]{babel}
\newcommand{\NMaterials}{775 }
\newcommand{\NMaterialsNoSpace}{775}

\newcommand{\NMaterialsHSE}{270 }

\begin{document}

    \title{Electronic properties of binary compounds with high fidelity and high throughput}
    
    \author{Protik Das}
    \thanks{mail-to: info@exabyte.io}
    \affiliation{Exabyte Inc., San Francisco, California 94103, USA}
    \affiliation{University of California, Riverside, California 92507, USA}
    \author{Timur Bazhirov}
    \thanks{PD and TB contributed equally to this work}
    \affiliation{Exabyte Inc., San Francisco, California 94103, USA}

    \begin{abstract}

    We present example applications of an approach to high-throughput first-principles calculations of the electronic properties of materials implemented within the Exabyte.io platform\cite{exabytePlatform, 2018-exabyte-accessible-CMD}. We deploy computational techniques based on the Density Functional Theory with both Generalized Gradient Approximation (GGA) and Hybrid Screened Exchange (HSE) in order to extract the electronic band gaps and band structures for a set of \NMaterials binary compounds. We find that for HSE, the average relative error fits within 22\%, whereas for GGA it is 49\%. We find the average calculation time on an up-to-date server centrally available from a public cloud provider to fit within 1.2 and 36 hours for GGA and HSE, respectively. The results and the associated data, including the materials and simulation workflows, are standardized and made available online in an accessible, repeatable and extensible setting.

\end{abstract}
    \maketitle
    \section{Introduction}
\label{sec:introduction}

    First-principles modeling for the purpose of materials design and discovery by means of high-throughput calculations received much attention with multiple success stories reported and multiple large-scale efforts deployed to date\cite{jain2013materialsproject, curtarolo2012aflowlib, saal2013openQMD, pizzi2016aiida, nomad}. Such efforts establish the data-centric approach to materials science and new collaborative inter-disciplinary work. The electronic properties of materials, such as the electronic band structures and band gaps, represent a natural target for the first principles modeling and have been successfully extracted earlier. Despite the attention, however, reaching high fidelity of the resulting predictions for the electronic structural properties is often given a lower priority during the high-throughput screening due to the complexity of the computational implementation of the underlying techniques.
    
    Among the initiatives that assembled large datasets of electronic properties of materials are the Materials Project \cite{jain2013materialsproject}, AFLOW\cite{curtarolo2012aflowlib}, the Open Quantum Materials Database\cite{saal2013openQMD}, AIIDA\cite{pizzi2016aiida} which also provides a set of building blocks for the construction of the simulation workflows, NOMAD\cite{nomad} with an open access data repository and data analytics tools, and the Computational 2D Materials Database\cite{haastrup2018computational, rasmussen2015computational}. All except the latter have primarily focused on obtaining results within the most robust, but less accurate implementation of Density Functional Theory within the Generalized Gradient Approximation, which has a number of well-known drawbacks when applied to the extraction of the electronic properties of materials\cite{perdew1996generalized}.
    
    We present the approach implemented in Exabyte.io\cite{exabytePlatform} and earlier described in more details in our prior publication \cite{2018-exabyte-accessible-CMD}. 
    The approach is able to deliver high fidelity in a repeatable way transferable from one material to another and able to facilitate high throughput as well. We apply this approach to extract the electronic properties of binary semiconducting compounds. We employ Density Functional Theory in the plane-wave pseudopotential formalism\cite{hohenberg-kohn1964DFT, mlcohen1979pseudopotentialDFT} and obtain the electronic band structures and band gaps for a diverse set of \NMaterials binary compounds, further called EBSC-\NMaterialsNoSpace. We extract the results within the Generalized Gradient Approximation \cite{perdew1996generalized} for the full set, and within the Hybrid Screened Exchange\cite{heyd2003hybrid} for a subset of \NMaterialsHSE materials. We compare the results with the available experimental data and present the assessment of the accuracy levels for each model. 
    
    This manuscript is structured as follows. We first explain the categorization of materials studied with respect to the modeling workflows and discuss the methodology and the parameters used using during the calculations. Next, we present the results for all the materials, compare them with the available experimental and computational data, and review the outliers. Finally, we discuss and analyze the results more in depth and suggest the pathways toward further improved accuracy. This work presents all the following: the results, the tools that generated the results, all associated data, and an easy-to-access way to reproduce and further improve upon our work.\cite{exabytePlatformBGPhaseIIIURL}

    %-------------------------------------------------------------------------------
%                                   SECTION
%-------------------------------------------------------------------------------
\section{Methodology}
\label{sec:methodology}

\subsection{General logic}

    We demonstrate the general execution flow employed in this work in a prior publication \cite{2018-exabyte-accessible-CMD}. As it is mentioned therein, our general routine includes the creation of the modeling workflows, followed by the ingestion of the structural data about materials and its conversion to database entries, followed by the execution of simulation jobs and further analysis of the results. In this work we employ the simulation workflows within the Generalized Gradient Approximation (GGA) and the Hybrid Screened Exchange (HSE) approach with a similar logic as previously explained. The users of Exabyte platform can clone the associated entities (eg. materials, workflows, jobs) - and re-create our calculations in order to reproduce or further improve the results.

\subsection{Materials}

    All materials studied in this work constitute the EBSC-\NMaterials set and are divided in 7 categories according to the difficulty levels of the corresponding simulation workflows. The details about the materials studied, including the corresponding counts per categories are given in Table \ref{table:difficulty}. Our selection is based initially on the binary semiconducting compounds (i.e. compounds with non-zero band gap) available from the Materials Project (MP) database\cite{jain2013materialsproject}. We optimized the choice of compounds according to the size of the crystal unit cell and the associated difficulty levels for the workflows as explained below. For the HSE calculations we used a subset containing \NMaterialsHSE total compounds.
    
    % \begin{figure}
    % \centering
    %     \includegraphics[width=0.48\textwidth]{figures/gga_spacegroup_plot.png}
    %     \label{fig:gga_sg}
    %     \includegraphics[width=0.48\textwidth]{figures/hse_spacegroup_plot.png}
    %     \label{fig:hse_sg}
    % \caption{Spacegroup distribution of materials for (a) GGA and (b) HSE calculations.}
    % \label{fig:sg_dist}
    % \end{figure}

\subsection{Workflows}

    In order to organize the information about the simulation workflows we employ the categorization illustrated in Table \ref{table:difficulty} and earlier explained in a prior publication \cite{2018-exabyte-accessible-CMD}. The categorization depends on: (a) the inclusion of the semi-core electronic states in the pseudopotentials, (b) the treatment of spin-orbit coupling within the calculation, and (c) the treatment of magnetic interactions. As it can be seen from the table, we prioritize compounds with lower difficulty levels where the computational burden is lower and focus our attention on the first 5 categories.

    \begin{table}[t!]
    	\centering
    	\begin{tabular}{c |  c   c  c | c c c}
    		\hline
    		\hline
    		Difficulty & Semi-core & SOC   & Magnetism & N$_{mat}^{GGA}$ & N$_{mat}^{HSE}$ & N$_{at}^{max}$ \\
    	    \hline
    	    1                  & no        & no    & no   & 159 & 129 & 24 \\
      	    2                  & yes       & no    & no   & 126 & 71  & 20 \\
      	    3                  & no        & yes   & no   & 303 & 43  & 20 \\
      	    4                  & yes       & yes   & no   & 123 & 8   & 20 \\
      	    5                  & no        & no    & yes  & 60  & 19  & 20 \\
      	    6                  & no        & no    & yes  & 2   & 0   & 16 \\
      	    7                  & yes       & yes   & yes  & 3   & 0   & 16 \\
    		\hline
    		\hline
    	\end{tabular}
    	\caption{
    	    Summary of the simulation workflows categorization employed in this work. "Semi-core" indicates that the pseudopotentials with semi-core states were used, "SOC" stands for the inclusion of the spin-orbit coupling, and "Magnetism" is used to denote the inclusion of collinear magnetic moments, except for the difficulty 7 when spin-orbit coupling and magnetism are included both, which lead to the treatment of non-collinear magnetic interactions. ``GGA" indicates number of materials calculated with GGA level of theory. Similarly ``HSE" indicates band gaps calculated with HSE level of theory.
    	}
    	\label{table:difficulty}
    \end{table}

\subsection{Computational setup}

    \subsubsection{Model and precision}

    All calculations were performed using Density Functional Theory\cite{kresse1996efficient, kohn1965self} in the plane-wave pseudopotential projector augmented wave (PAW)\cite{blochl1994projector} formalism using the Vienna Ab initio Simulation Package (VASP)\cite{kresse1996software, hacene2012accelerating}. Within the generalized gradient approximation the exchange-correlation effects were modeled using the Perdew-Berke-Ernzerhof (PBE)\cite{perdew1996generalized} functional, for hybrid screened exchange the Heyd-Scuseria-Ernzerhof (HSE06) implementation\cite{HSE06Reference} is used. We sample in the reciprocal cell based on the k-points per reciprocal atom (KPPRA) with a uniform unshifted grid. The KPPRA value of 2,000 is used. The band structures are calculated using the default paths per each crystal lattice type as defined in the AFLOW methodology\cite{curtarolo2012aflowlib}. The electronic density of states calculations were performed on a denser grid with KPPRA of 16,000 using tetrahedron interpolation as implemented in VASP\cite{kresse1996software}.
    
    \subsubsection{Method and parameters used}
    
    Atomic positions for all structures studied were optimized until the forces acting on each atom are less than $0.01$ eV/\AA. The kinetic energy cutoff was set to 520 eV. We used the Gaussian smearing algorithm, and the blocked Davidson iteration scheme\cite{johnson2001block} as the initial minimization algorithm for the calculation of the electronic subsystem, with the ionic positions updated using the conjugated gradient algorithm. A smearing value of $50$ meV was employed. The semi-empirical Grimme-D2 correction was used\cite{grimme2006vdWcorrection}. The HSE calculations were performed with the default 25$\%$ mixing parameter for the short-range Hartree-Fock exchange\cite{heyd2003hybrid}. The screening parameter $\mu$ is set to 0.2 \AA$^{-1}$. 
    
    \subsubsection{Hardware and related}
    
    We performed all calculations described in this work using the high-performance computing hardware readily available from Microsoft Azure cloud computing service\cite{azure-instance-types} in the same manner as described in a prior publication\cite{2018-exabyte-accessible-CMD} using H16r and H16mr instances. Computational resources were provisioned and assembled on-demand by the software available within the Exabyte platform\cite{exabytePlatform}. The calculations were executed within a two-week period from June 6th to June 20th of 2018. The peak size of the computational infrastructure used during this work was administratively limited to 750 nodes or over 10,000 total computing cores. In this manuscript we omit the discussion of the infrastructure capable of supporting the large-scale calculations deployed during this work, however, we note that this work serves as another proof point demonstrating the viability and readiness of cloud computing for the high-performance computing workloads involved in the first-principles computational materials modeling\cite{exabyte2018hp3c}.

\subsection{Data access and repeatability}

    The data about the materials, workflows, batch jobs for each material, and the associated properties, including the files for each step of the simulation workflows are all available online.\cite{exabytePlatformBGPhaseIIIURL, exabytePlatformBGPhaseIIIData}. Readers may freely access the results. Alternatively, in order to reproduce the results, the readers may create an account, copy materials and/or workflows to their account collection, and recreate the simulations for these materials. Optionally, the readers may also adjust the model parameters, and improve upon our results. For further assistance on how to do that the readers may consult the online documentation\cite{exabyteDocumentation}.

    \section{Results}
\label{sec:results}
    
\begin{figure}[ht!]
        \label{fig:all}
        \centering
        \includegraphics[width=0.48\textwidth]{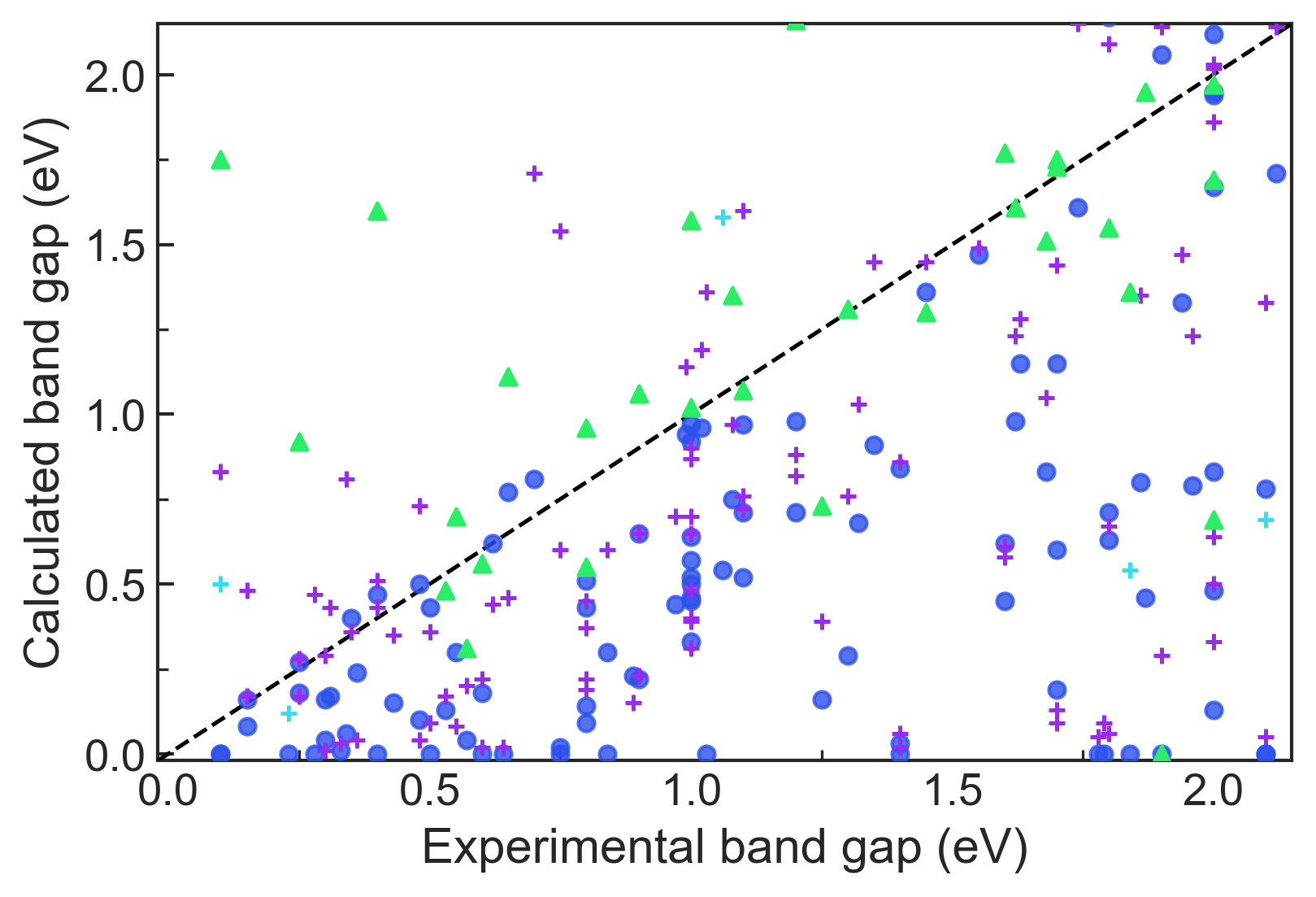}
        \includegraphics[width=0.48\textwidth]{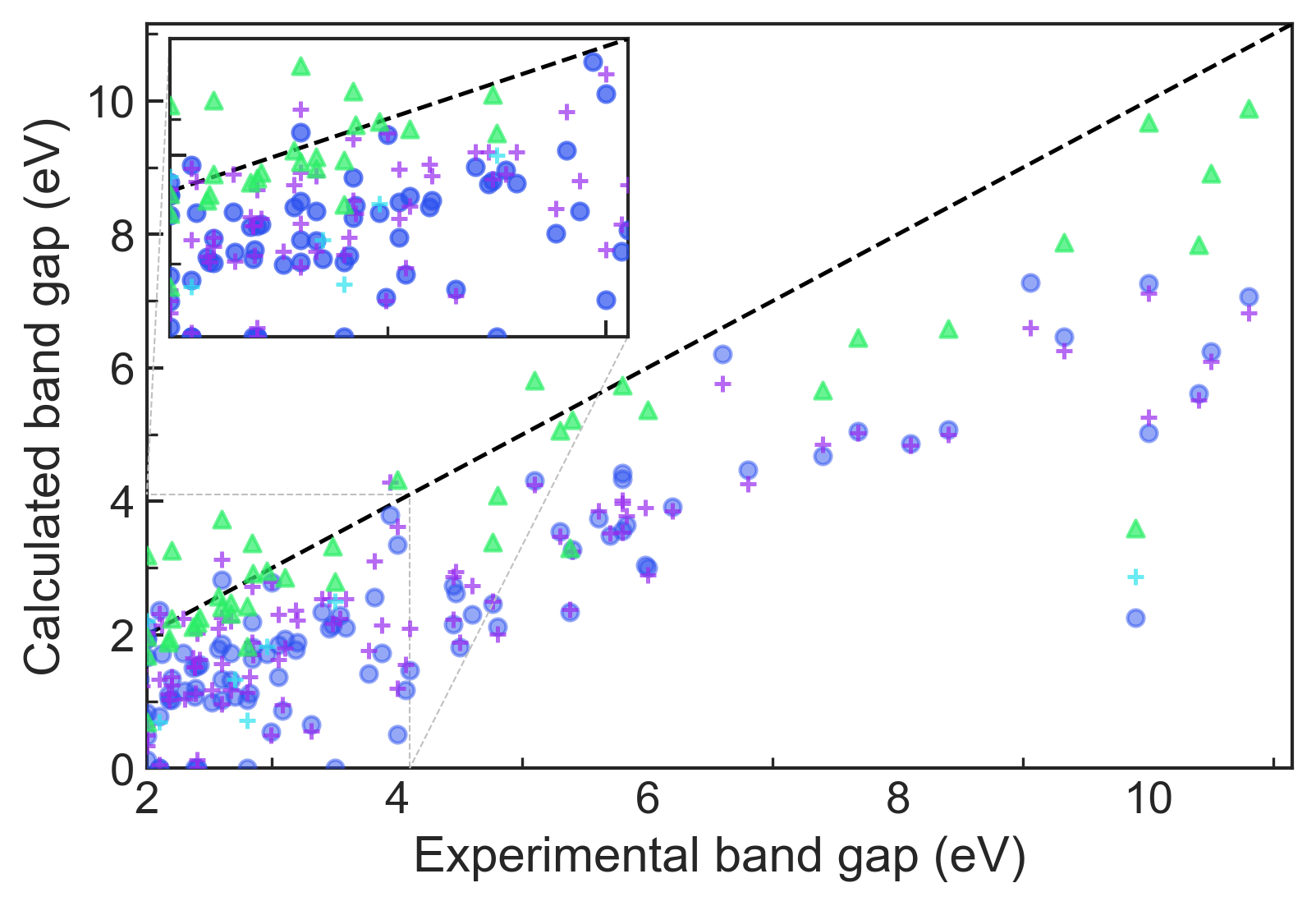}
        \caption{
            Comparative plot of the calculated and experimentally available values for the electronic band gaps obtained in the current work. Legend: GGA and HSE denote the results of this work for the corresponding level of theory. MP-GGA and MP-GGA+U denote the results of Materials Project\cite{jain2013materialsproject} available at the moment of this writing and calculated within the GGA and GGA+U approaches, respectively. The legend is the same as Fig. \ref{fig-d1}. We separate the materials with the experimental gaps in the 0-2 eV range into the top sub-figure. The compounds with gaps larger that 2 eV are shown in the bottom sub-figure. The latter also has an inset where the 2-4 eV region is expanded for better visibility.
    }
    \end{figure}

\begin{figure}[ht!]
        \includegraphics[width = 0.48\textwidth]{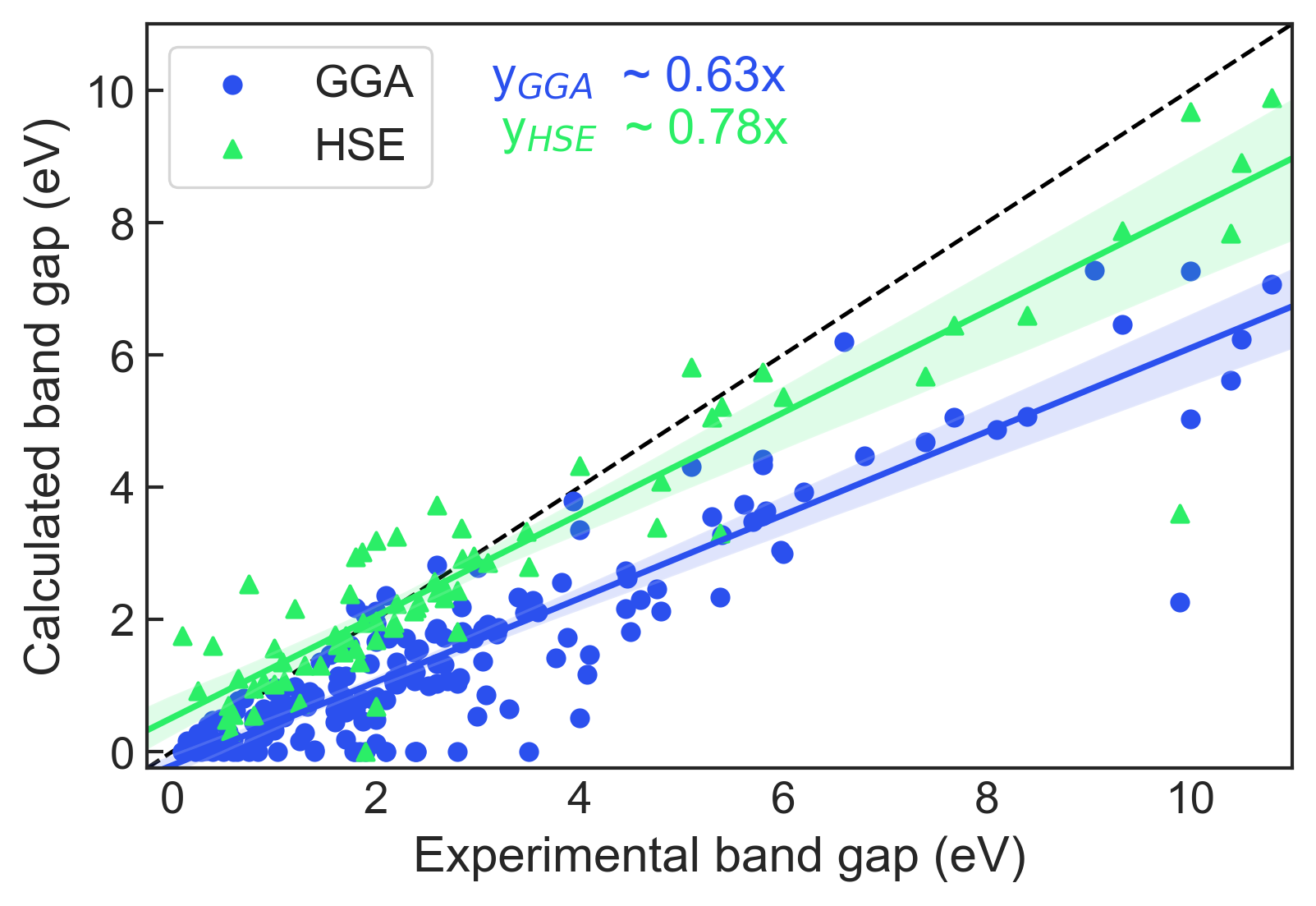}
        \caption{
            Comparative plot of the calculated and experimentally available values for the electronic band gaps obtained in the current work We include a linear $y = kx + b$ fit to data per each model (GGA, HSE). The legend is same as in Fig.\ref{fig:all}. The proportionality coefficients for each of the linear fits are shown in the figure.
        }
        \label{fig-reg}
    \end{figure}

    We present the comparison of all calculated band gaps within GGA and HSE with their experimental values\cite{strehlow1973compilation, oxygenVacancies2012deml} for the materials where the experimental data is available in Fig. \ref{fig:all}. We also include the results of Materials Project\cite{jain2013materialsproject} (further referred to as MP) calculated within GGA and GGA+U approaches for reference (further referred to as MP-GGA and MP-GGA+U, respectively). We organize and make all data available online for the readers at this link\cite{exabytePlatformBGPhaseIIIData}. As expected, the GGA underestimates the band gaps, and HSE significantly improves the results. We fit the results with a linear regression fit, as shown in Fig. \ref{fig-reg}. It can be seen that for a simple $y = kx + b$ relation the resulting values for the model-wise errors based on the coefficient of proportionality $k$ are: GGA - 37\%, HSE - 22\%. We proceed to discuss the results per each difficulty level below and focus on the "outlier" cases where the results deviate significantly from their expected values.
    
    % ------------------------------------------------------- %
    %                   Figure 1
    % ------------------------------------------------------- %   
    
    % ------------------------------------------------------- %
    %                   SUBSECTION
    % ------------------------------------------------------- %
    \subsection{Difficulty 1 (D1)}
    \label{subsec:d1}
    
\begin{figure}[ht!]
        \includegraphics[width = 0.48\textwidth]{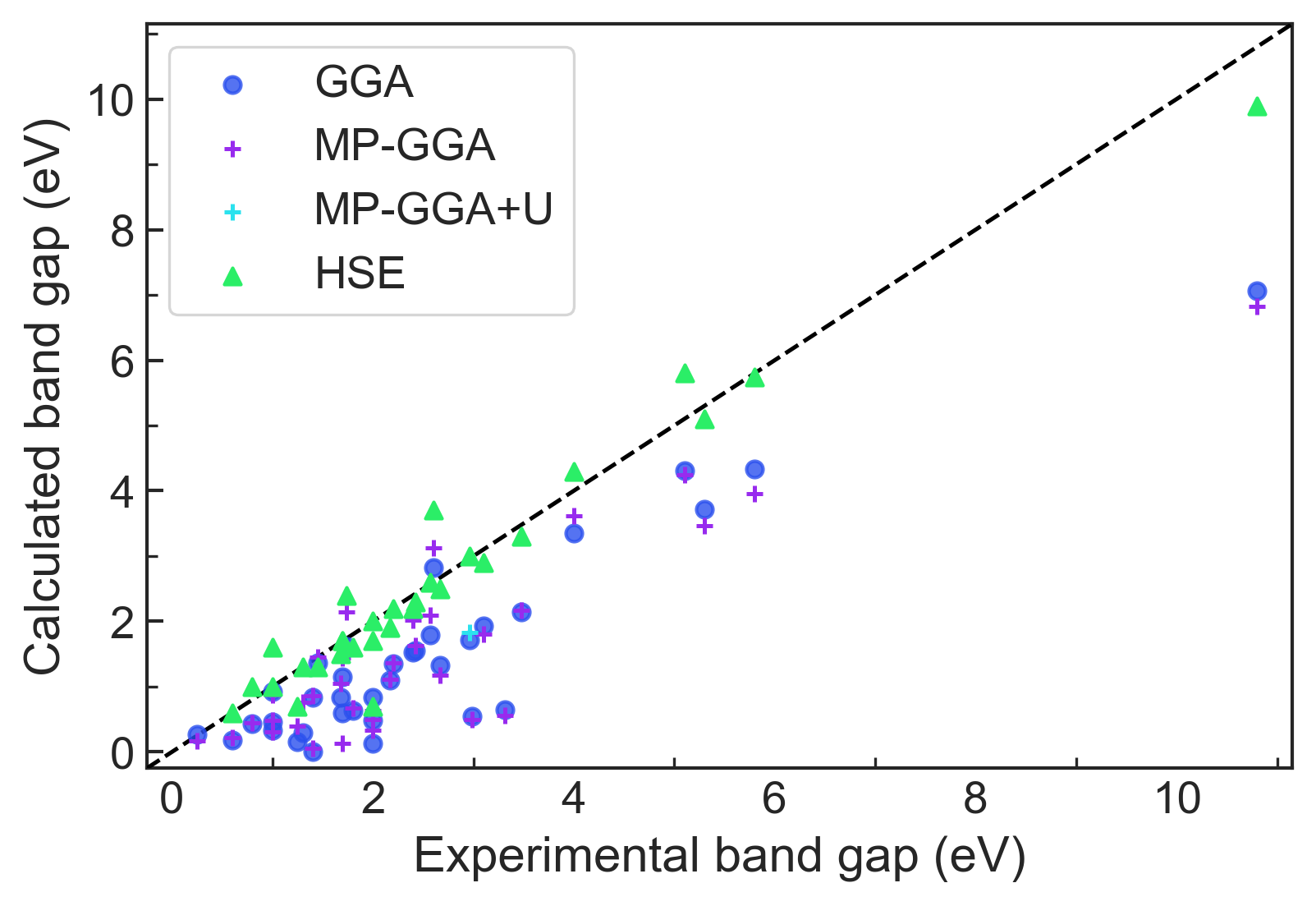}
        \caption{
            Comparative plot of the calculated and experimentally available band gap values for the materials in the D1 category.
        }
        \label{fig-d1}
    \end{figure}

        Fig. \ref{fig-d1} shows the band gaps of materials for difficulty level 1. Overall, our results are in good agreement with MP for this difficulty level, with minor differences related to the inclusion of semi-core pseudopotentials in our case. The average error for GGA calculations is 46.2\%. HSE calculations improve the band gap and reduce the average error to 13\%. For MgF$_2$, with the experimentally found gap of 10.8 eV\cite{strehlow1973compilation} is an outlier case. Both GGA and MP-GGA underestimate the value by 34.5\% and 36.8\%, respectively, and HSE reduces the band gap error to 8\%. At the same time, the lattice constant of MgF$_2$ matches within 0.04\% of the experimental values\cite{vidal1979neutron}. 
        
    % ------------------------------------------------------- %
    %                   SUBSECTION 
    % ------------------------------------------------------- %
    \subsection{Difficulty 2 (D2)}
    \label{subsec:d2}
    
\begin{figure}[ht!]
        \includegraphics[width = 0.48\textwidth]{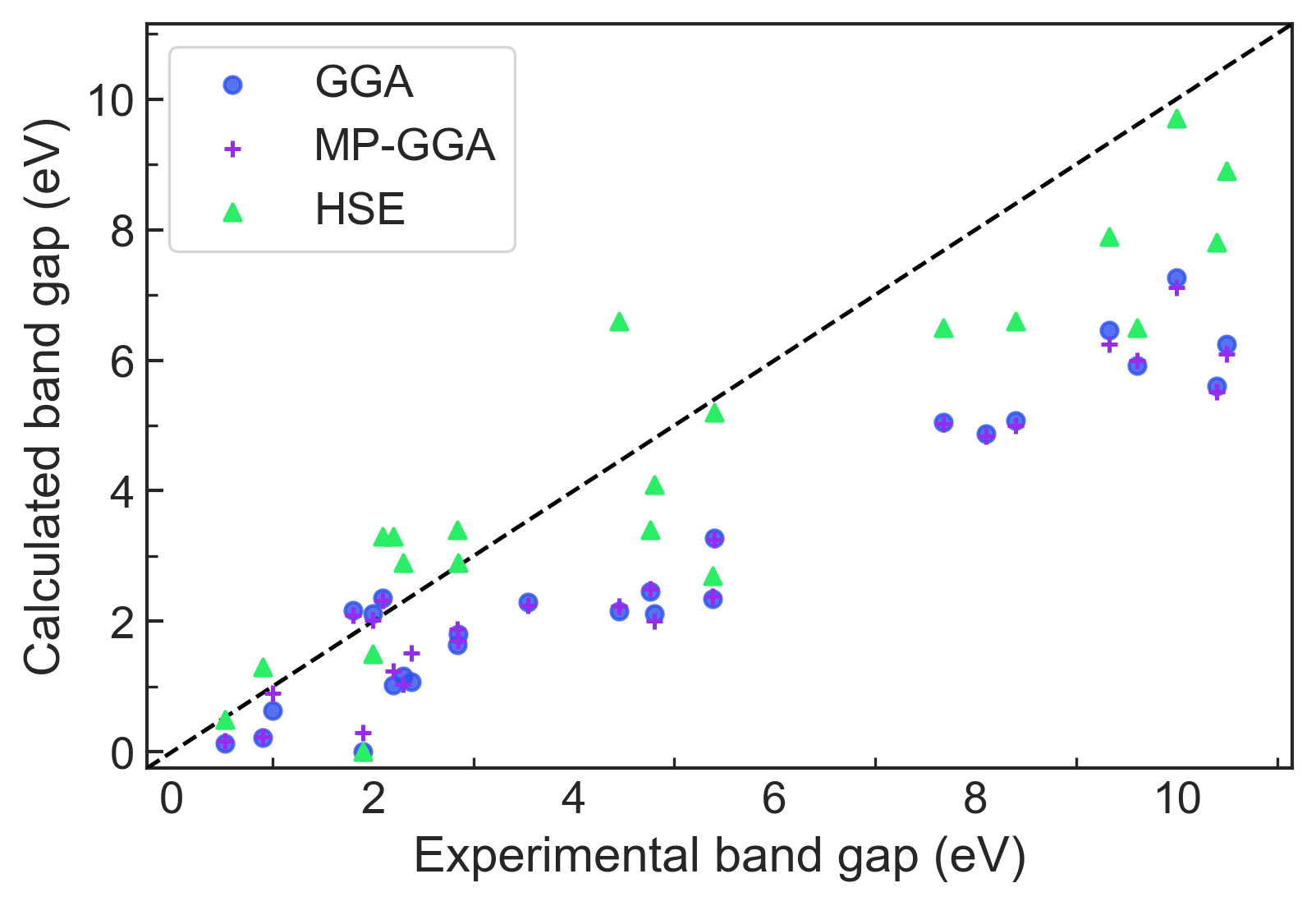}
        \caption{
            Same as Fig. \ref{fig-d1} for the D2.
        }
        \label{fig-d2}
    \end{figure}

        Fig. \ref{fig-d2} compares the calculated band gaps of D2 materials with their experimental values. Here we see good agreement with MP results as well. Within GGA the average error is 43.4\%. For HSE the average error is reduced to 27.6\%. SrSe is an important outlier case in the figure. The experimental gap of SrSe is 4.45 eV\cite{strehlow1973compilation}. The lattice constants of the relaxed structure match experiment within 0.1\% \cite{primak1948x}. Both GGA and MP-GGA underestimate the band gap by 51.5\% and 48.3\%, respectively. On the contrary, HSE calculation with the default 25\% exact exchange overestimates the band gap by 48.3\%. This suggests that the mixing parameter needs to be adjusted in order to predict the gap correctly.
        
        For Ca$_2$Si, both GGA and HSE calculations predict zero band gap (with a semi-metallic nature of the band structure) while MP predicts a band gap of 0.29 eV. We include semi-core electronic states for both elements in our calculations (titled "pv" for Ca and "sv" for Si), MP calculations do so only for Ca (titled "sv"). In addition, the lattice constants $b$ and $c$ of our relaxed structure are 9.9\% and 3.4\% smaller compared to the MP result. All of this contribute to the reduced value of the band gap in our case.

    % ------------------------------------------------------- %
    %                   SUBSECTION
    % ------------------------------------------------------- %
    
    \subsection{Difficulty 3 (D3)}
    \label{subsec:d3}
    
\begin{figure}[ht!]
        \includegraphics[width = 0.48\textwidth]{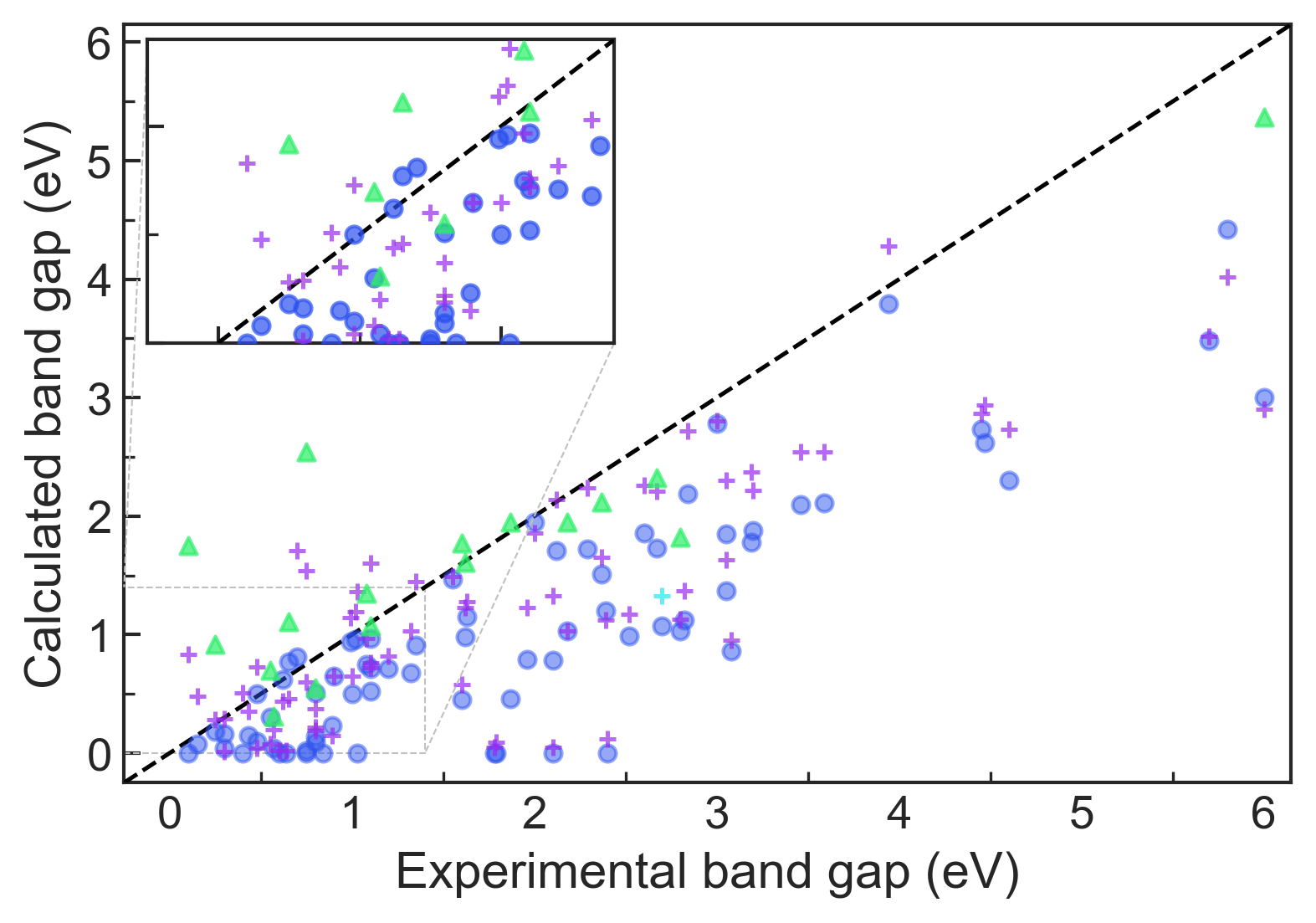}
        \caption{
            Same as Fig. \ref{fig-d1} for the D3. Legend follows Fig. \ref{fig-d1}.
        }
        \label{fig-d3}
    \end{figure}

        The results for the D3 category are shown in Fig. \ref{fig-d3}. For most of the materials the GGA band gaps agree well with the MP-GGA values. We concentrate on discussing the outliers below.
        
        PtS$_2$ and PtSe$_2$ are layered two-dimensional (L2D) materials with experimental band gap values of 0.75 eV and 0.10 eV, respectively. The MP-GGA band gaps for PtS$_2$ and PtSe$_2$ are overestimated by 105.7\% and 727.4\% compared to the experimental values. As the interaction between the layers of L2D materials is dominated by the van-der-Waals (vdW) forces, the inclusion of vdW correction is necessary to calculate the structural parameters accurately. It is our understanding that due to the absence of vdW correction within MP, the lattice constant $c$ is overestimated by 24\% and 20.4\%\cite{furuseth1965redetermined}, respectively, which leads to the overly large band gaps. Our GGA calculations predict PtS$_2$ and PtSe$_2$ to be metallic. The lattice constant $c$ within our calculation is underestimated compared to the experimental value by 13.5\% and 8.56\% only, which we believe closes the band gaps in both. We suspect that the DFT-D2 correction incorporated in our calculation does not fully capture the vdW interaction in these materials\cite{zhao2016extraordinarily, zhang2017mechanism}.
        
        HgS has an experimental gap of 0.70 eV\cite{strehlow1973compilation}. GGA and MP-GGA both overestimate it by 16.1\% and 143.8\%, respectively. The lattice constant $a$ within GGA is 8.9\% smaller than MP-GGA value which can account for the band gap difference. Ag$_2$S has an experimental band gap of 1.03 eV. Our calculations predict Ag$_2$S to be metallic and MP-GGA predicts a band gap of 1.36 eV. The kinetic energy cutoff parameters used in our work are same as in MP calculations. The final relaxed structure in our case, however, is different than that of MP, perhaps due to the inclusion of spin-orbit coupling and vdW correction. The final total energy of the structure is 0.29 eV/atom lower in our case, which might indicate a different stable phase.
        
        Pr$_2$O$_3$ and CdSe have experimental band gaps of 0.84 eV and 1.87 eV\cite{strehlow1973compilation}, respectively. MP-GGA overestimate the values by 345.9\% and 92\%. While MP used pseudopoential titled "Pr 3" which has 11 electronic states in valence, we used the default pseudopotential (titled "Pr default") with 13 states in valence and smaller cutoff radius. We include a further discussion on this in section \ref{sec:discussion}). The band gap within MP is calculated on a coarse grid of k-points which might also contribute to the error. For CdSe, the lattice constant in the calculation matches the experiment within 1\% \cite{landolt1987111}. MP shows the band gap as 3.58 eV while the band structure plot suggests a gap value of about 0.5 eV at the moment of this writing.
        
        Te$_2$Ru, P$_2$Pd and As$_2$Pt have experimental band gaps of 0.25 eV, 0.65 eV and 0.55 eV, respectively. HSE predictions for these materials are 0.92 eV, 1.11 eV and 0.70 eV, and hence are overestimated by 267.3\%, 70.6\% and 27.3\%. It appears that in order to get accurate gap values for these materials within HSE, the tuning of the exact exchange mixing parameter is needed.

    % ------------------------------------------------------- %
    %                   SUBSECTION
    % ------------------------------------------------------- %
    
    \subsection{Difficulty 4 (D4)}
    \label{subsec:d4}
    
\begin{figure}[ht!]
        \includegraphics[width = 0.48\textwidth]{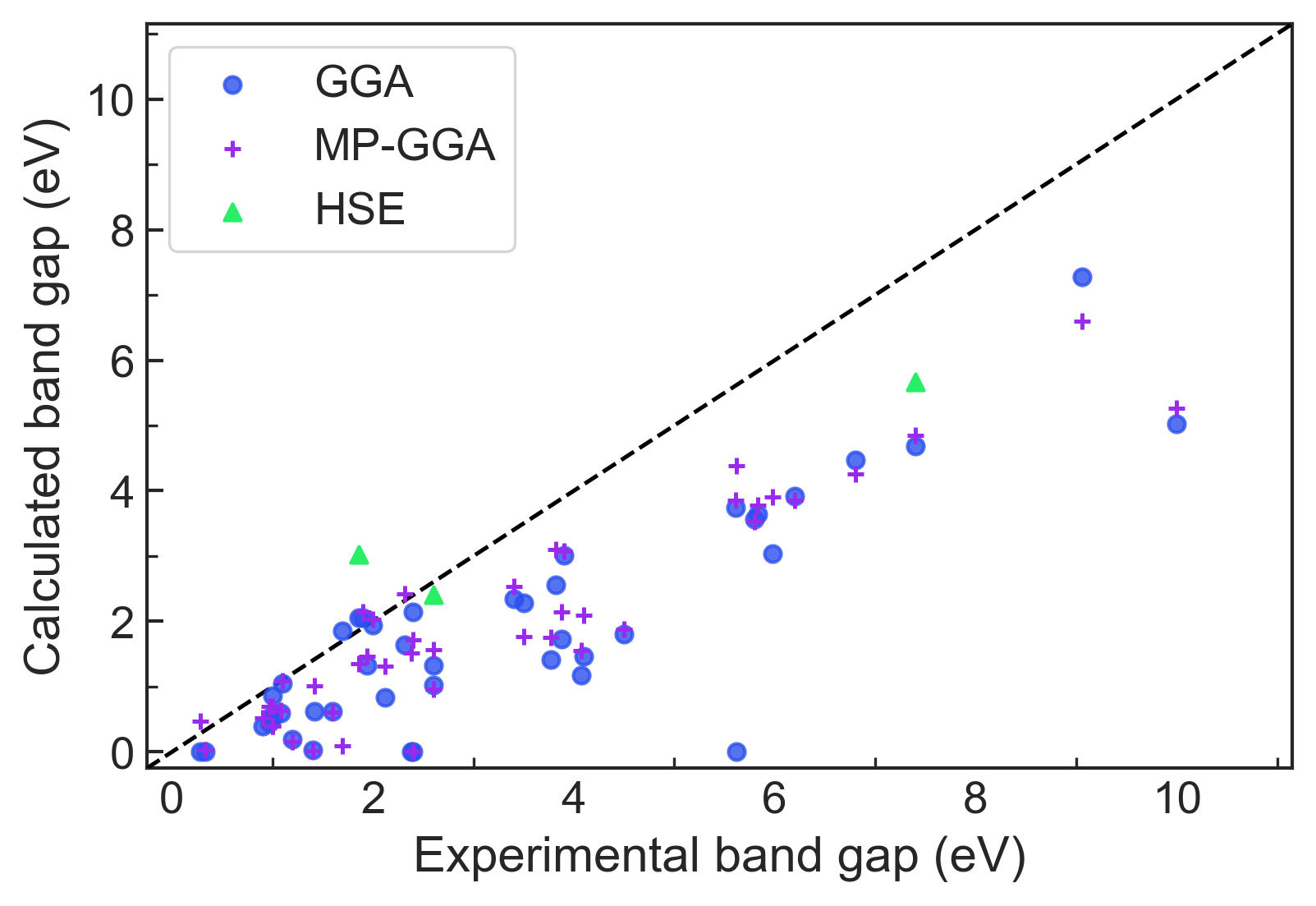}
        \caption{
            Same as Fig. \ref{fig-d1} for the D4.
        }
        \label{fig-d4}
    \end{figure}

        Fig. \ref{fig-d4} compares the calculated band gaps with their experimental values for the D4 category. Our results agree well with that of MP except for the following cases. PbS has an experimentaly found gap of 0.28 eV. MP-GGA result for this material is 0.47 eV, whereas our GGA calculation predicts PbS to be semi-metallic with no bands crossing. The lattice constant of the relaxed structure of PbS is 1.5\% smaller compared to MP-GGA which might lead to no gap in our case. CsF has an experimental gap of 10 eV. The band gap predicted by GGA and MP-GGA are 5.03 eV and 5.26 eV, respectively which are significantly underestimated compared to the experimental value by 49.7\% and 47.4\%. 

    % ------------------------------------------------------- %
    %                   SUBSECTION
    % ------------------------------------------------------- %
    
    \subsection{Difficulty 5 (D5)}
    \label{subsec:d5}
    
\begin{figure}[ht!]
        \includegraphics[width = 0.48\textwidth]{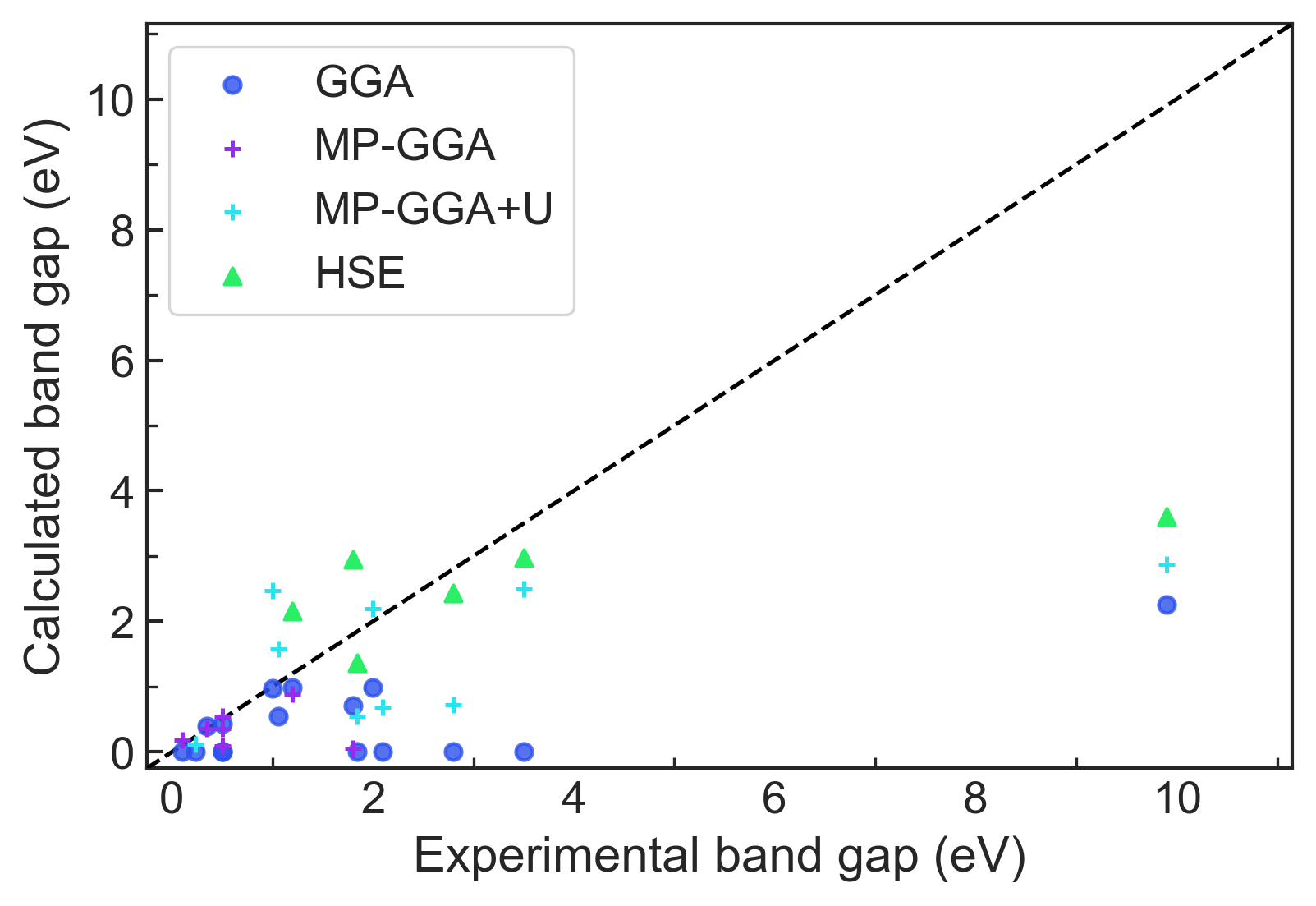}
        \caption{
            Same as Fig. \ref{fig-d1} for the D5.
        }
        \label{fig-d5}
    \end{figure}

        Fig. \ref{fig-d5} has the band gaps for category D5 materials. FeO and CoO have experimental band gaps of 2.10 eV and 2.8 eV\cite{deml2015intrinsic}, correspondingly. In our GGA calculations, these materials are predicted to be metallic. Inclusion of GGA+U within MP leads to non-zero band gaps. For CoO, for example, the MP-GGA+U result leads to an error of 74\%. Within our HSE calculation, the error is reduced to 13.2\%. 
        
        For MnF$_2$ the experimental band gap is 9.90 eV\cite{strehlow1973compilation}. Our GGA calculation predicts 2.26 eV with an error of 77.2\%. The MP-GGA+U value is close to ours at 2.87 eV. HSE produces the estimate of 3.6 eV, still with a 63\% error. The large difference between the experimental and calculated band gaps indicate that the experimental value might be extracted for a different polymorph of MnF$_2$. For instance, MP has three polymorphs of this material and the values for the energy above Hull are within 47 meV of each other for all.
        
    \section{Discussion}
\label{sec:discussion}
    
    We meant this study as a data-centric benchmark of the ability of the current generation of pseudopotential density functional theory (DFT) to predict the electronic properties of materials. We also focused our attention on how it can be applied in an accessible way with minimal additional computational setup (i.e. no specialized hardware or compilation routines). We elaborate on the results of our prior work\cite{2018-exabyte-accessible-CMD} and report the results for a significantly larger number of compounds.
    
    \begin{figure*}[ht!]
        \label{fig:heatmap}
        \includegraphics[width=\textwidth, height=4.4in]{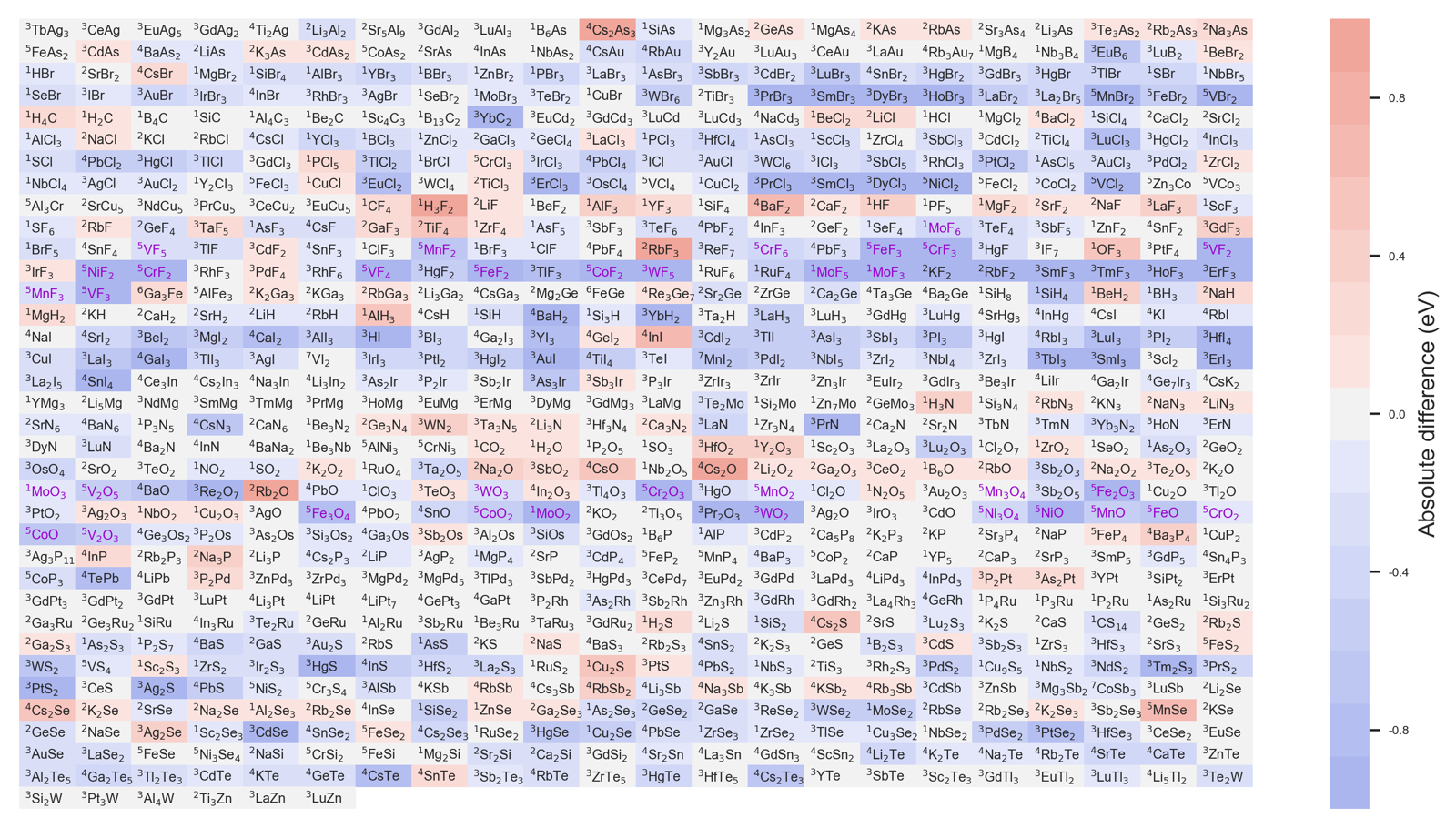}
        \includegraphics[width=\textwidth, height=4.4in]{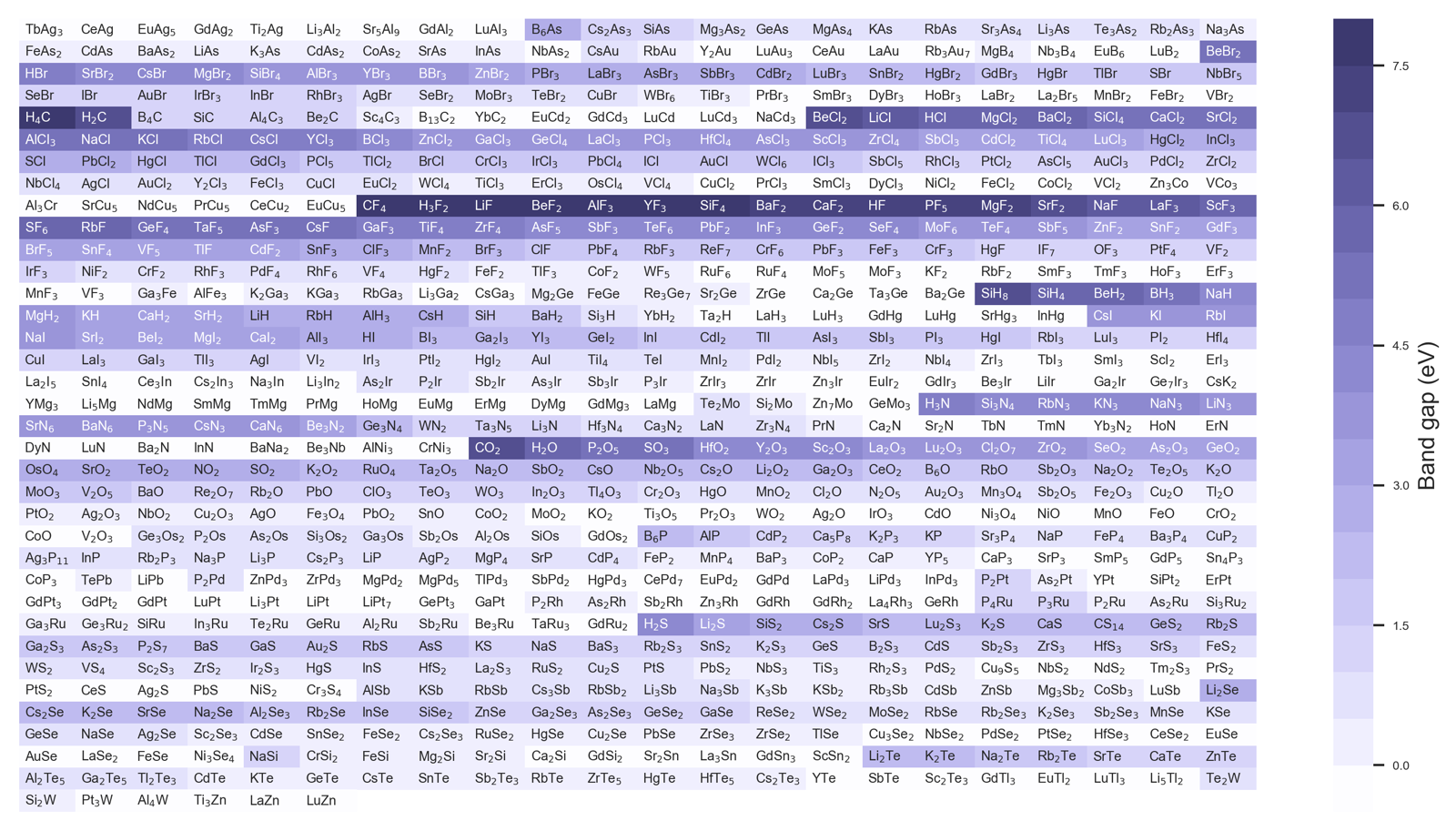}
        \caption{
            (TOP) Heatmap plot for $E_{abs}^{BG}$, the absolute difference in the band gap values calculated within GGA in this work and MP\cite{jain2013materialsproject}, as defined in Eq. \ref{eq:abs}. Chemical formulas for compounds where MP results use GGA+U are highlighted in violet. (BOTTOM) Same for the GGA band gap results of this work. Both plots use same compounds order sorted by the second element in formula and the gap value starting from the top right corner. Superscripts denote the corresponding difficulty levels as defined in the methodolody section.
        }
    \end{figure*}
    
    % ------------------------------------------------------- %
    %                   SUBSECTION: Fidelity
    % ------------------------------------------------------- %
    
    \begin{figure}[ht!]
        \includegraphics[width = 0.48\textwidth]{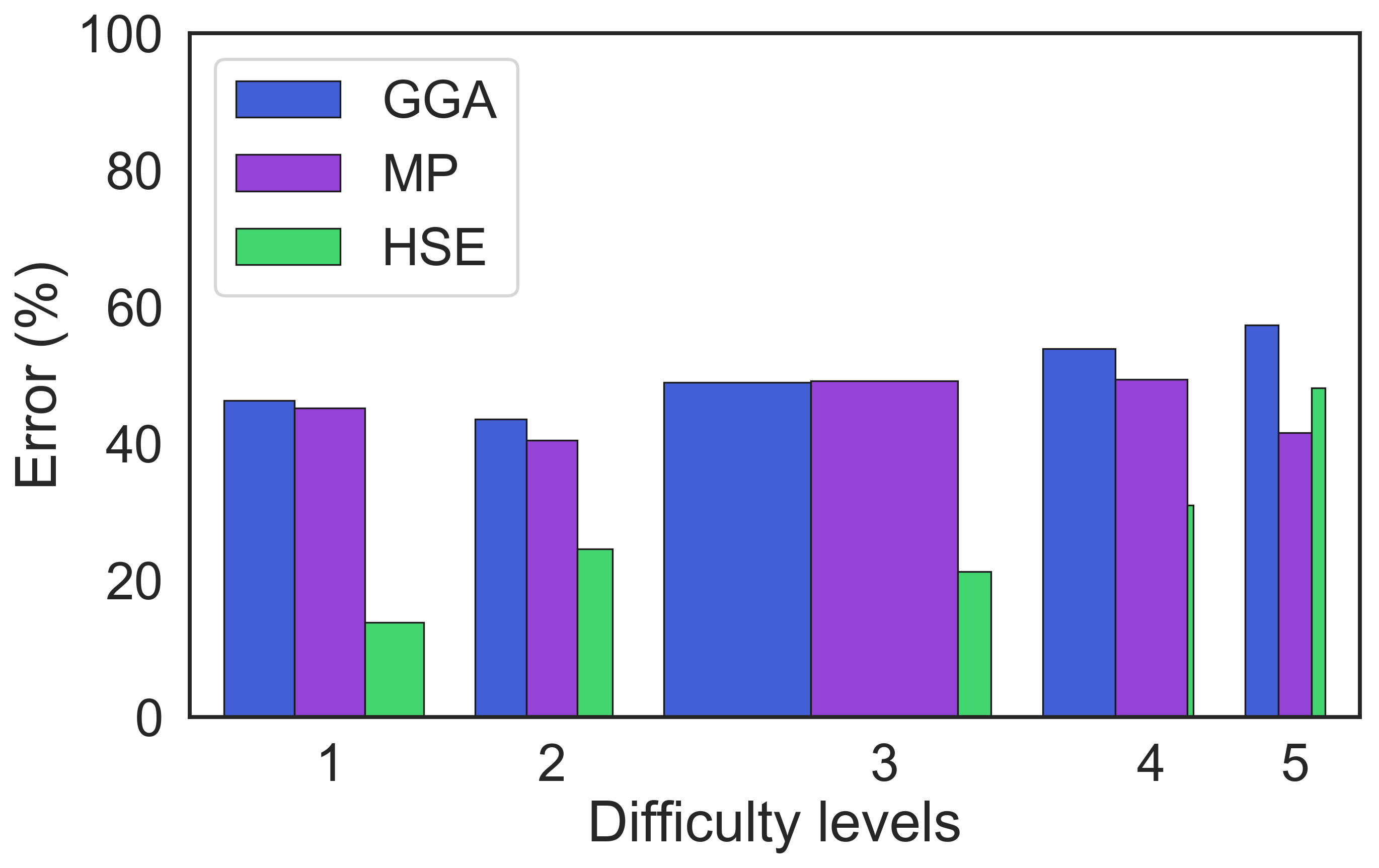}
        \caption{
            Difficulty-wise average errors. The width of the bars are proportional to the number of materials in category.
        }
        \label{figure:difficulty-wise-error}
    \end{figure}    
    
    \subsection{Fidelity and error analysis}
    \label{subsec:fidelity}
    
        \subsubsection{Comparison with experimental results}
    
    When comparing with the available experimental data we point out some important conditions used within our approach that are known to affect the calculation results. Similar to our prior study\cite{2018-exabyte-accessible-CMD}, we introduce a van-der-Waals (vdW) correction as implemented in VASP\cite{kresse2010vdWCorrection, grimme2006vdWcorrection}, and conduct the structural relaxation within the GGA and subsequently use the resulting structure for HSE calculations. Figure \ref{figure:difficulty-wise-error} has the data about the average errors per each category. The width of each column is proportional to the number or materials.
    
    Unlike the prior study, where we manually assembled the experimental results, here we programmatically collect the data using an online platform\cite{citrine}. At the moment of this writing this platform only contains the stoichiometric data and lacks the structural parameters enough to uniquely identify the materials. We therefore apply filters and remove from consideration some data points as described further. We start with 231 total experimental data entries\cite{strehlow1973compilation}. We completely remove from comparison the 23 entries where GGA results are over 25\% larger than the experimental values. In addition, when calculating the averages, we remove the 2 cases where the HSE values are over 150\% larger than experimental: Sb$_2$Os (0.2 eV), Te$_2$Ru (0.25 eV). The resulting average values for the errors are 49\% and 22\% for GGA and HSE over the sets of 215 and 75 total experimental entries, correspondingly.
    
        \subsubsection{Comparison with MP data}
    
    We conduct a comparison with the results of MP for all calculated values within GGA (and exclude the MP-GGA+U). The top and bottom parts of Fig. \ref{fig:heatmap} illustrate the absolute and relative band gap differences. The absolute difference $E_{abs}^{BG}$ is defined as:
    
    \begin{equation}
        \label{eq:abs}
        E_{abs}^{BG} = sign(E_{bg}-E_{MP}) \times min(E_{bg}-E_{MP}, 1.0),
    \end{equation}
    
    where we intentionally limit the maximum difference to 1.0 eV for the visualization purposes. $E_{bg}$ is the band gap calculated in this work, $E_{MP}$ is the value from MP, and $sign$ and $min$ are the sign and minimum functions.
    
    As can be seen from $E_{abs}^{BG}$, generally our values match closely with MP-GGA gaps. The cases when the values are different were discussed in the previous section and are generally attribute to one of the following: (a) the difference in pseudopotentials used, eg. inclusion of the semi-core states in this work, which leads to the somewhat better GGA results for band gaps (this is very evident for Ge, for example, in our prior study\cite{2018-exabyte-accessible-CMD}); (b) inclusion of the van-der-Waals correction in our calculations, which leads to the shortened structural parameters, and, therefore, smaller band gaps. Due to the inclusion of vdW correction and spin-orbit coupling the relaxed structures in our calculations often have smaller lattice constants compared to that of MP-GGA.
    
    The gap value correction also occurs for materials containing lanthanides Nd, Pr, Eu, Gd, Ho, Dy. For these compounds our calculations predict metallicity whereas MP suggests the presence of a gap. Such a discrepancy leads to the relative difference of 100\%. It appears as a common practice to use pseudopotentials with a reduced number of states in valence for the elements in the Ce-Lu row for VASP calculations in order to reduce the computation time, and MP is following this convention, which results in non-zero band gaps. In our calculations the default pseudopotentials are used where all $f$ electrons are considered, and this fact is leading to the metallic behavior. We demonstrated it clearly for the HoN, for example (and included the comparison in the data available online): our calculations reproduce the MP gap value of 0.087 eV (also notably rather small compared to the smearing employed for the electronic occupations) when a "Ho\_3" pseudopotential is used, for the "default" Ho pseudopotential we get no gap.
    
    Another important note can be derived from KF$_2$ and RbF$_2$ that are extracted as metals in our calculation, although both are semiconducting in MP results. The Fermi level for both compounds is calculated to be close to the top of the conduction band, however, and we believe that both compounds should indeed be perceived as wide-gap semiconductors with a gap value in the 5-7 eV range. MP results suggest values below 2 eV and, perhaps erroneously, ferromagnetic ordering in these compounds. We attribute the difference in the gap values to the absence of magnetic treatment in our case, and to the pseudopotentials used for alkali metals with "p" and "s/p" states in valence in our and MP case, respectively.
    
    % ------------------------------------------------------- %
    %                   SUBSECTION: Improvements
    % ------------------------------------------------------- %
    \subsection{Further improvements to accuracy}
    \label{subsec:improvements}
    
    One way to improve the accuracy of the HSE calculations would be to use an adjustment scheme where first the improved value for the mixing parameter is calculated per material based on a statistical model, and then a single HSE calculation is executed. This work suggests that within the HSE the band gaps are significantly overestimated for the following materials (experimental values in braces): Sb$_2$Os (0.2 eV), Te$_2$Ru (0.25 eV), As$_2$Pt (0.55 eV), P$_2$Pd (0.65 eV), TiS$_3$ (1.30 eV), FeP$_2$ (1.60 eV), Mg$_3$As$_2$ (1.60 eV), SiSe$_2$ (2.40 eV), and SrSe (4.45 eV). Finding a robust way to classify the outliers with respect to a set of descriptors based on the atomistic data would provide the next step in building such a scheme.

    Alternatively, another way to improve the accuracy of the results would be to use a dynamically self-consistently adjustable value for the HSE mixing parameter similar to how it is done in \cite{scHSE2014galli}. This approach would be more computationally intensive as it requires the convergence of the static dielectric constant with respect to the mixing parameter to be achieved during the calculation. As we demonstrated earlier\cite{2018-exabyte-accessible-CMD}, the GW approximation, both in the G0W0 and in the self-consistent implementation can be used efficiently in order to further improve the accuracy of band gap predictions.
    
    \begin{figure}[hb!]
        \label{figure:difficulty-wise-runtime}
        \includegraphics[width = 0.48\textwidth]{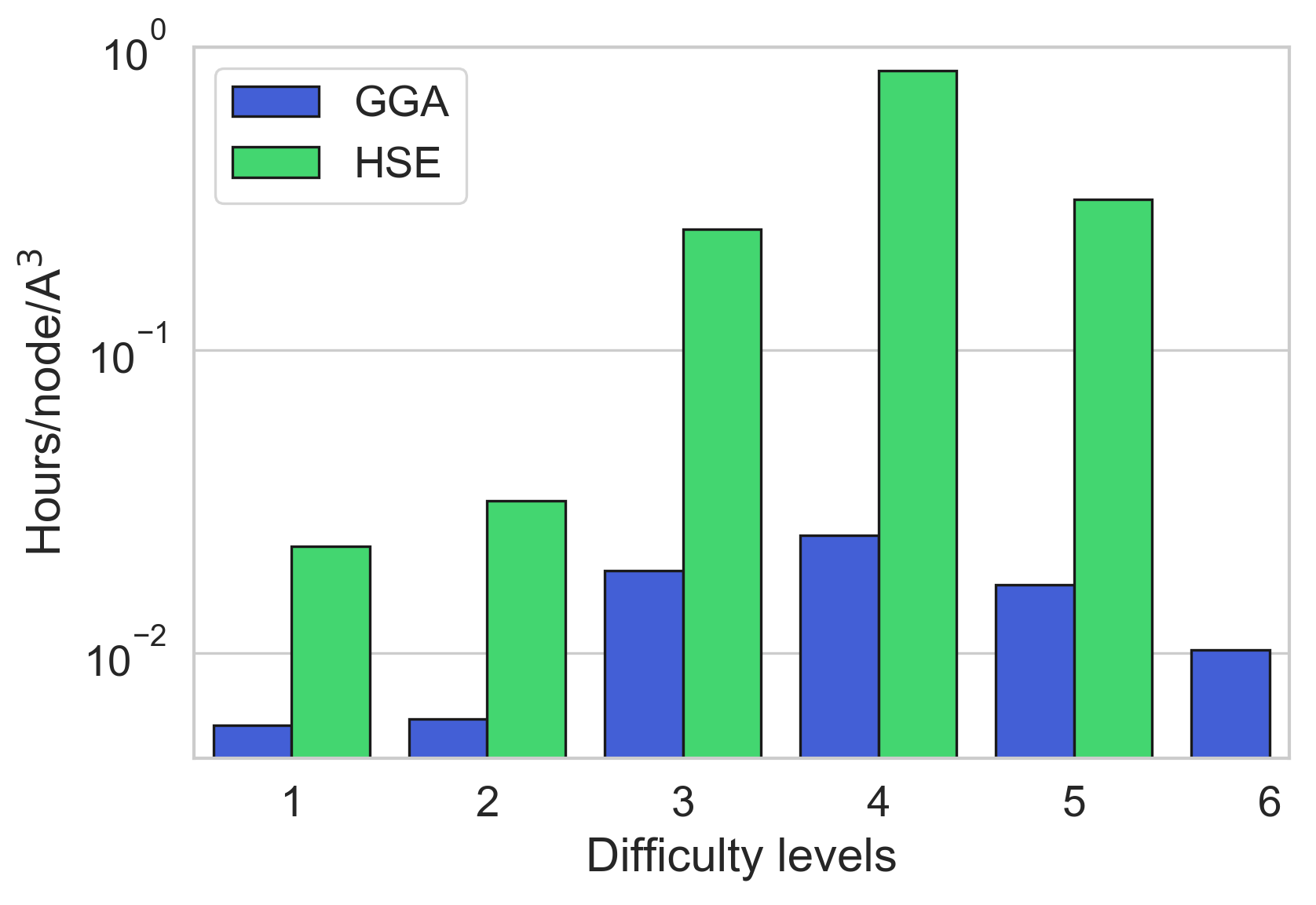}
        \caption{
            Calculation time per each difficulty level (as defined in section \ref{sec:methodology}). The time is normalized per one compute node and unit cell volume (\AA$^3$).
        }
    \end{figure}

    % ------------------------------------------------------- %
    %                   SUBSECTION: Costs
    % ------------------------------------------------------- %
    \subsection{Computational time and cost}
    
    Figure \ref{figure:difficulty-wise-error} has the data about the average calculation runtime per each category for both GGA and HSE. Similar to our prior consideration from \cite{2018-exabyte-accessible-CMD}, in order to provide an insight into the feasibility of improved accuracy approaches, we construct a simple logarithmic regression using the data obtained for the GGA and the HSE results. We assume that the average simulation lifetime increases exponentially as the average error is dropping. As can be seen from Table \ref{table:exact-accuracy}, within this logic one would need to run a simulation for about 30 days on average in order to produce an exact result.
    
    Our motivation for the above is to provide a metric of the extent to which the physics-based first-principles modeling can augment the trial-and-error experimental approach when compared with respect to the capital and time investments required. We suggest that for the equivalent of one month of calculation time (human time) on a commodity compute server readily available from a cloud provider it is possible to obtain results that are accurate well within 20$\%$ and potentially within 1-5$\%$ range for the properties that we study in the current work.

    \begin{table}[ht!]
        \label{table:exact-accuracy}
    	\centering
    	\begin{tabular}{l |  c   l  c | l}
    		\hline
    		\hline
    		Model          & Avg. Err, ($\%$) & Avg. time & Cost ($\$$) & Note \\
    	    \hline
    	    \hline
    	    *Exact*        & 0        & 30 days   & 5,000   & extrapolated   \\
      	    HSE            & 24       & 35 hrs    & 200     & factual   \\
      	    GGA            & 46       & 1.2 hrs   & 10      & factual   \\
    		\hline
    		\hline
    	\end{tabular}
    	\caption{
    	    Average errors and the associated average calculation time (human time) for the HSE and GGA cases studied in this work. *Exact* and *Zero* values are constructed through a simple logarithmic fit of the HSE/GGA data for the (hypothetic) models that would produce exact and zero-fidelity results correspondingly. NOTE: only the subset of materials where HSE values are also available is used to calculate the average error for GGA case shown above.
    	    }
    \end{table} 

    % ------------------------------------------------------- %
    %                   SUBSECTION: Future outlook
    % ------------------------------------------------------- %
    \subsection{Future outlook}

    Computational materials design is rapidly evolving toward a data-driven science where the modeling results are aggregated and classified by their precision/accuracy, as the recent work from NOMAD demonstrates\cite{nomad}. Major improvements in the way computational materials science is used would be possible when increased veracity of this data also becomes a norm. The approach described in this work can assist with achieving this goal.
    
    Modeling workflows accessible in a standardized and repeatable way let the field evolve away from the "medieval artisan-like" model\cite{pizzi2016aiida} still prevalent nowadays. This work serves as a proof that precision within 20\%, perhaps only for the electronic materials at this moment, is readily achievable using existing first-principles modeling techniques. 
    
    Our intent is to welcome collaborative contributions in order to, firstly, further grow the online repository of high fidelity results; secondly, allow contributions from other high-fidelity modeling techniques beyond studied here; and, finally, facilitate the creation of statistical (machine learning) models based on the available data.

    \section{Conclusions}
\label{sec:conclusions}
    
    We present the applications of a novel approach to materials modeling from nanoscale implemented within the Exabyte platform\cite{exabytePlatform} and capable of delivering both high fidelity and high throughput in an accessible and data-centric manner to a set of \NMaterials binary semiconducting compounds. We report the results for the electronic band gaps obtained within the Generalized Gradient Approximation (GGA) for the full dataset and with Hybrid Screened Exchange (HSE) for a subset of \NMaterialsHSE materials. We analyze the level of fidelity for the predictions delivered by each of the models used, compare the results with experimental data and prior similar calculation attempts, when available, and discuss the corresponding computational costs and pathways to further improved accuracy.
    
    We find the average relative error in the estimates for the electronic band gaps obtained in this work to be 22$\%$ for HSE and 49$\%$ for GGA, respectively. We further find the average calculation time on a current up-to-date compute server centrally available from a public cloud provider to fit within 1.2 and 36 hours for GGA and HSE, correspondingly. We present not only the results and the associated data, but also an easy-to-access way to reproduce and extend the results by means of the Exabyte platform.\cite{exabytePlatformBGPhaseIIIURL} Our work provides an accessible and repeatable practical recipe for performing high-fidelity first-principles calculations of the electronic structural properties of materials in a high-throughput manner.
    
    \bibliography{references}

\end{document}